\documentclass[aip,sd,amsmath,amssymb,preprint]{revtex4-1}

\usepackage{graphicx}
\usepackage{dcolumn}
\usepackage{bm}
\usepackage{subcaption}

\usepackage[english]{babel}

\begin{document}

\title{Pulse length of ultracold electron bunches extracted from a laser cooled gas}

\author{J.G.H. Franssen}
\affiliation{Department of Applied Physics, Eindhoven University of Technology, P.O. Box 513, 5600 MB Eindhoven, The Netherlands}
\affiliation{Institute for Complex Molecular Systems, Eindhoven University of Technology, P.O. Box 513, 5600 MB Eindhoven, The Netherlands}
\author{T.L.I. Frankort}
\affiliation{Department of Applied Physics, Eindhoven University of Technology, P.O. Box 513, 5600 MB Eindhoven, The Netherlands}
\author{E.J.D. Vredenbregt}
\affiliation{Department of Applied Physics, Eindhoven University of Technology, P.O. Box 513, 5600 MB Eindhoven, The Netherlands}
\affiliation{Institute for Complex Molecular Systems, Eindhoven University of Technology, P.O. Box 513, 5600 MB Eindhoven, The Netherlands}
\author{O.J. Luiten}
\email{o.j.luiten@tue.nl}
\affiliation{Department of Applied Physics, Eindhoven University of Technology, P.O. Box 513, 5600 MB Eindhoven, The Netherlands}
\affiliation{Institute for Complex Molecular Systems, Eindhoven University of Technology, P.O. Box 513, 5600 MB Eindhoven, The Netherlands}

\date{\today}

\begin{abstract}

We present measurements of the pulse length of ultracold electron bunches generated by near-threshold two-photon photoionization of a laser-cooled gas. The pulse length has been measured using a resonant $3$~GHz deflecting cavity in TM$_{110}$ mode.
We have measured the pulse length in three ionization regimes. The first is direct two-photon photoionization using only a $480$~nm femtosecond laser pulse, which results in  short ($\sim 15~$ps) but hot ($\sim 10^{4}~$K) electron bunches.
The second regime is just-above-threshold femtosecond photoionization employing the combination of a continuous-wave $780$~nm excitation laser and a tunable $480$~nm femtosecond ionization laser which results in \emph{both} ultracold ($\sim 10~$K) \emph{and} ultrafast ($\sim 25~$ps) electron bunches. These pulses typically contain $\sim 10^{3}$ electrons and have an rms normalized transverse beam emittance of $1.5\pm0.1~$nm$\cdot$rad. The measured pulse lengths are limited by the energy spread associated with the longitudinal size of the ionization volume, as expected. 
The third regime is just-below-threshold ionization which produces Rydberg states which slowly ionize on microsecond time scales.

\end{abstract}

\pacs{37.10, 37.20, 41.75, 41.85}

\maketitle

\section{Introduction}

Ultrafast electron diffraction has developed into a powerful technique for studying structural dynamics\cite{King2005,Sciaini2011a,Zewail2010,Vanacore2015}. Pumping samples with femtosecond laser pulses and probing them with high energy electron bunches can easily lead to sample damage, which is particularly true for biological molecules\cite{Egerton2015}. This means that diffraction patterns preferably have to be captured with a single electron bunch: single-shot electron diffraction. This requires electron bunches with $10^{6}-10^{7}$ electrons\cite{Egerton2015,Engelen2014} per pulse which are prone to space charge explosions, resulting in loss of temporal resolution and degradation of transverse beam quality. The effect of space charge forces may be mitigated by using an ultracold electron source since it allows for larger source sizes and thus lower bunch densities for the same beam quality\cite{Engelen,McCulloch2013}.

To obtain high quality diffraction images the relative energy spread of the electron bunch needs to be much smaller than unity and the transverse coherence length larger than the lattice spacing of the structure under investigation. Previous work demonstrated that the ultracold electron source is capable of producing high quality diffraction images\cite{Engelen,VanMourik2014a,Speirs2015a}, even with electron bunches created by femtosecond photoionization\cite{Engelen,VanMourik2014a}. 
The transverse beam quality of the ultracold electron source has been investigated extensively previously\cite{Engelen2013,Engelen2014,VanMourik2014a}. However, the longitudinal electron beam characteristics have not been investigated in great detail. Recently it was shown that the ultracold electron source can be used to create ultracold electron bunches with a root-mean-square (rms) pulse length of 250 ps\cite{Sparkes2016a}. In this paper we present pulse length measurements with sub-ps resolution of ultracold electron bunches with an rms pulse duration of $25$ ps containing $~\sim 10^{3}$ electrons per pulse. Up to $\sim 10^{6}$ electrons can be extracted in a single shot from the ultracold electron source\cite{Speirs2015a} but then strong space charge effects come into play. These space charge effects can be minimized by shaping the initial electron distribution~\cite{Luiten2004,Thompson2016}. In this paper we stick to maximally $\sim 10^{3}$ electrons per bunch, thus avoiding the complication of space charge effects.

For UED, the electron pulse length has to be shorter than the shortest timescale associated with the process under investigation. This means that often the electron bunch length should preferably be much shorter than one picosecond. This can be achieved by compressing longer electron pulses using a resonant RF cavity\cite{VanOudheusden2010b} in TM$_{010}$ mode, but for this to work the electron pulse should not be longer than a few $10$~ps\cite{VanOudheusden2010b} for a cavity operated at $3~$GHz. In this paper we show that such electron bunches can indeed be extracted from the ultracold electron source.

This paper is organized as follows: In Section~\ref{secultracoldelec} we introduce the ultracold electron source and the relevant photoionization schemes (Section~\ref{secultracoldelec}A) that were used. In Section~\ref{secultracoldelec}B and \ref{secultracoldelec}C we address the transverse electron beam quality (transverse emittance) and the longitudinal beam quality (longitudinal emittance). In Section~\ref{secultracoldelec}D we discuss a model of the rubidium ionization process allowing us to make an estimate of the shortest electron pulse lengths that can be expected. In Section~\ref{sectionexperi} the experimental setup is described. Finally, in Section~\ref{secresultsanddisc}, the results will be presented.
We show that we can produce electron bunches which are \emph{both} ultracold \emph{and} ultrafast, with rms pulse durations of $\sim 20$ ps, short enough to be compressed to sub-ps bunch lengths. In Section~\ref{outlookconc} we will finish with a conclusion and an outlook.

\section{Ultracold electron source}\label{secultracoldelec}

\begin{figure}[htb!]
\centering	
\includegraphics[width=15cm]{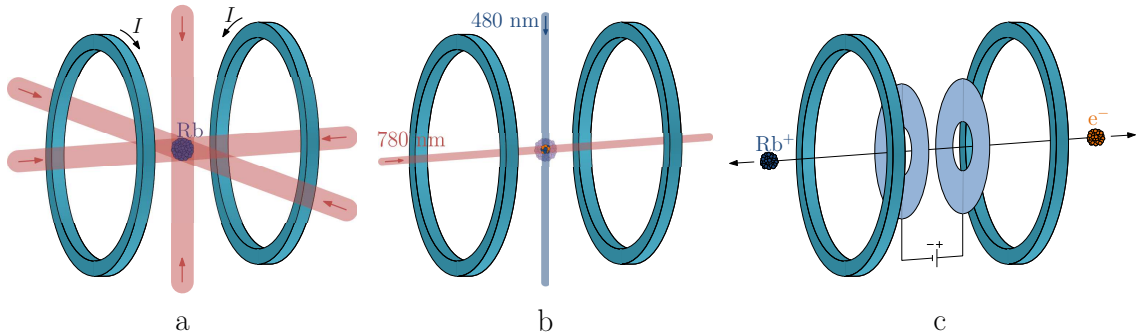}
\caption{\label{mot}Schematic representation of the electron bunch production sequence. (a) Laser-cooling and trapping of $^{85}$Rb in a MOT; indicated are six cooling laser beams (red) and two magnetic field coils in anti-Helmholtz configuration. (b) $708~$nm excitation laser and $480~$nm ionization laser intersecting at right angles, creating an ionized gas. (c) Acceleration of the resulting electrons and ions in the static electric field.}
\end{figure}

Ultracold electron bunches are created by near-threshold photoionization of laser-cooled and trapped 85-rubidium atoms, as is illustrated in Fig.~\ref{mot}. First, Fig.~\ref{mot}a, rubidium atoms are laser-cooled and trapped in a magneto-optical trap (MOT), with an atom density $\sim 10^{16}~$m$^{-3}$.

Second, Fig.~\ref{mot}b, the trapping laser is switched off $1~\mu$s before the ionization laser pulse, so all atoms in the MOT relax to the ground state. Then the $780$~nm excitation laser beam is switched on, creating a small cylinder of excited atoms in the $5P_{\frac{3}{2}} F=4$ state which has a rms radius of $35~\mu$m. Subsequently a small volume of rubidium atoms is ionized by a femtosecond $480$~nm ionization laser beam, intersecting the excitation laser beam at right angles, resulting in a cloud of cold electrons and ions typically less than $100~\mu$m in diameter. This all occurs in a static electric field\cite{Taban2008,Engelen2014,Engelen2013}, see Fig.~\ref{mot}c, which immediately accelerates all charged particles created.

\begin{figure}[htb!]
\centering	
\includegraphics[width=12cm]{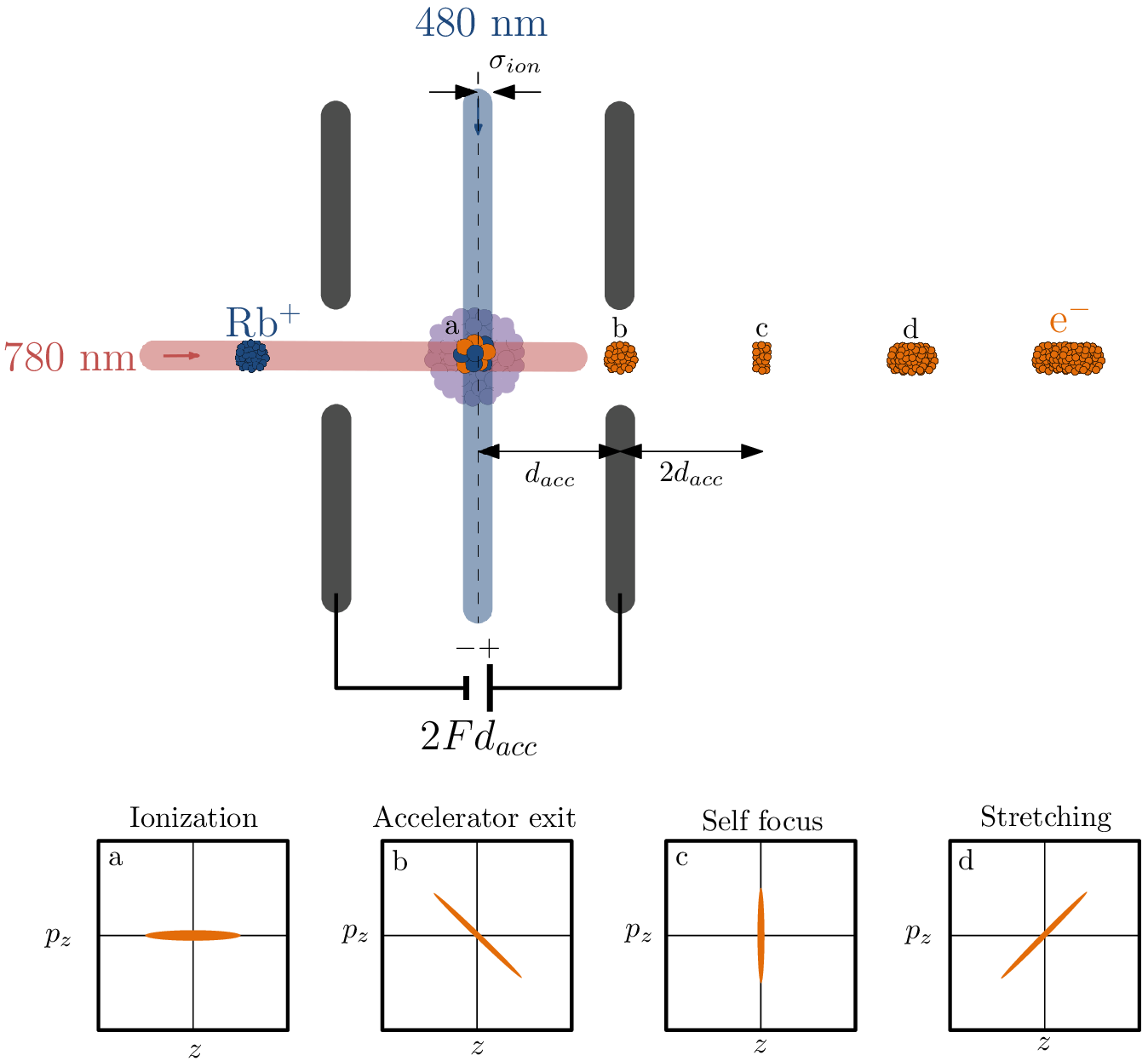}
\caption{\label{accelerator}Schematic representation of the acceleration of the ultra cold electron bunch. The electrons are accelerated over a distance $d_{acc}$ and the ionization laser has a width $\sigma_{ion}$. The bottom figures show the evolution of the longitudinal phase space distribution at various points in the beam-line. The electrons receive a correlated momentum spread upon exiting the accelerator structure (b), which causes the electron beam to first self focus (c) and then expand (d) as it drifts towards the detector.}
\end{figure}

Figure~\ref{accelerator} shows a schematic picture of the accelerator. The electrons are created in the center of the accelerating structure\cite{Taban2008} which has a potential $2 F d_{acc}$ applied across the electrodes, with $F$ the electric field strength and $d_{acc}=13.5~$mm the length over which the electrons are accelerated. Electrons created in the back of the ionization volume will be accelerated over a longer distance and thus acquire more energy than electrons created in the front. This results in a correlated momentum spread at the exit of the accelerator (Fig.~\ref{accelerator}b). At a distance $2d_{acc}$ behind the accelerator the electrons created in the back catch up with the electrons created in the front, creating a self focus (Fig.~\ref{accelerator}c). After this self focus the electron pulse stretches due to the energy spread (Fig.~\ref{accelerator}d).

In absence of space charge forces, the initial longitudinal energy spread is dominated by the width of the ionization laser beam $\sigma_{ion}$ in the direction of the acceleration field. The initial energy spread due to thermal motion is $1-10$~meV for all three degrees of freedom\cite{Sciaini2011a,Engelen,Engelen2013}, which is at least an order of magnitude smaller than the energy spread due to the finite ionization laser beam size, typically $0.1-1$~eV. Therefore the Boersch effect can be neglected. As a result, the rms relative energy spread $\frac{\sigma_{U}}{U}$ is to a good approximation directly proportional to the ionization laser beam size,

\begin{equation}\frac{\sigma_{U}}{U}=\frac{\sigma_{ion}}{d_{acc}}\label{relenergylasersize},\end{equation}

with $U=eFd_{acc}$ the average bunch energy. The spot size of the diffraction limited ionization laser beam focus at the position of the ionization volume results in $\sigma_{ion} \approx 30~\mu$m so the expected rms relative energy spread $\frac{\sigma_{U}}{U} \gtrsim 3\cdot10^{-3}$.

Depending on the intensity and the central wavelength of the $480~$nm femtosecond laser ionization may occur by different mechanisms. We may distinguish three different ionization schemes, schematically indicated in Fig.~\ref{ionschemerb}, which are described in more detail below. In principle all three ionization processes always occur but their relative importance is determined by the intensity and central wavelength of the ionization laser.

\subsection{Ionization schemes}

\begin{figure}[htbp]
\centering
\includegraphics[width=14cm]{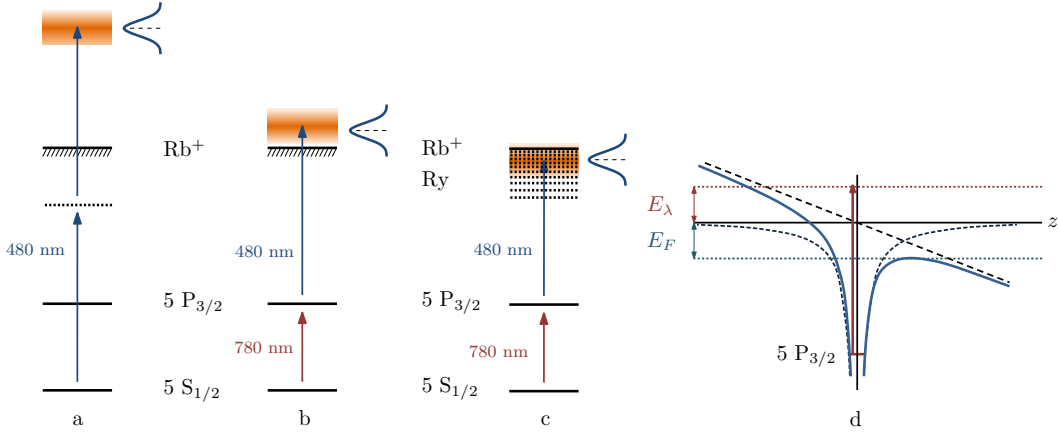}
\caption{\label{ionschemerb}Ionization schemes. (a) Photoionization from the rubidium ground state using two $480~$nm photons. (b) Just-above-threshold photoionization, using a $780~$nm and a $480~$nm photon. (c) Just-below-threshold photoionization; Rydberg states are created which slowly ionize. (d) Rubidium potential in a static electric field, with $E_{F}$ the contribution to excess energy due to the stark shift and $E_{\lambda}$ the contribution to excess energy due to the ionization laser wavelength.}
\end{figure}

The first ionization regime is photoionization of rubidium atoms from the ground state ($5S_{1/2}$) using two blue photons ($480-480$), see Fig.~\ref{ionschemerb}a. This will result in hot electrons ($\sim 10^{4}$~K) due to large excess energies. The ionization probability of this non-linear process scales with the intensity of the ionization laser field squared, which therefore results in an ionization volume that is smaller in length by a factor $1/\sqrt{2}$ than it would be by $780-480$~nm photoionization, described below. 

The second regime is near-threshold photoionization using a red and a blue laser ($780-480$), which results in ultracold electron bunches suitable for high quality electron diffraction. Atoms are excited from the ground state to the $5P_{3/2}$ state using a $1~\mu$s excitation pulse. In the mean time, an ultrafast blue laser ionizes the excited atoms, see Fig.~\ref{ionschemerb}b. In all ($780-480$) measurements the intensity of the femtosecond ionization laser has been decreased to minimize the contribution of two-photon ($480-480$) ionization. The ionization probability per $480~$nm photon of this process is at least an order of magnitude larger than that in the direct photoionization regime, resulting in more electrons per pulse.

When the center wavelength of the ionization laser is tuned below the ionization threshold, slowly decaying Rydberg states are formed ($780-480-$Ry), the third regime. For a given ionization laser wavelength $\lambda$ and electric field strength $F$, the excess energy\cite{} is given by:

\begin{equation}E_{exc}=E_{\lambda}+E_{F}\equiv hc\left(\frac{1}{\lambda}-\frac{1}{\lambda_{0}}\right)+2E_{h}\sqrt{\frac{F}{F_{0}}}\label{excessenergy},\end{equation}

where $\lambda_{0}=479.06$~nm is the zero-field ionization laser wavelength threshold, $E_{h}=27.2$~eV the Hartree energy, $F_{0}=5.14 \cdot 10^{11}$~V/m the atomic unit of electric field strength, $h$ Planck's constant and $c$ the speed of light. 

The part of the femtosecond laser spectrum lying in the positive excess energy regime will still be able to produce fast electron pulses, see Fig.~\ref{ionschemerb}c. Figure~\ref{ionschemerb}d depicts the Stark-shifted rubidium potential, with $E_{\lambda}$ the contribution to the excess energy due to the laser wavelength and $E_{F}$ the contribution to the excess energy due to the Stark shift.

The tail below threshold will create long lived Rydberg states that slowly ionize. We have scanned the center wavelength of the ionization laser across the zero excess energy point to investigate the ionization dynamics of the Rydberg gas.


\subsection{Transverse phase space}\label{waistscansecti}

The transverse beam quality of an electron beam can be described by the normalized root-mean-squared (rms) transverse emittance\cite{Geer2009a}

\begin{equation}\hat{\epsilon}_{x}=\frac{\sigma_{x}\sigma_{p_{x}}}{mc}=\sigma_{x}\sqrt{\frac{k_{b}T_{x}}{m c^{2}}}\label{emittancetrans}\end{equation}

with $\sigma_{x}$ the rms source size, $m$ the electron mass and $\sigma_{p_{x}}$ the rms momentum spread at the source, which can be expressed in terms of an effective transverse source temperature $T_{x}$.

The beam quality of the electrons extracted from the ultracold source has been investigated extensively in previous work\cite{Engelen2013,Engelen,Engelen2014,VanMourik2014a}. Waist scan measurements resulted in a normalized rms transverse beam emittance of $\hat{\epsilon}_{x}= 1.5$~nm$\cdot$rad\cite{Engelen2013,VanMourik2014a}. This is equivalent to a relative transverse coherence length of $C_{\perp} = \frac{L_{\perp}}{\sigma_{x}}= \frac{\lambdabar_{c}}{\hat{\epsilon}_{x}}= 2.5 \cdot 10^{-4}$ with $\lambdabar_{c}\equiv\frac{\hbar}{mc}$ the reduced Compton wavelength, $\hbar$ Dirac's constant and $L_{\perp}$ the transverse coherence length. 
\begin{figure}[htb!]
\centering
\includegraphics[width=10cm]{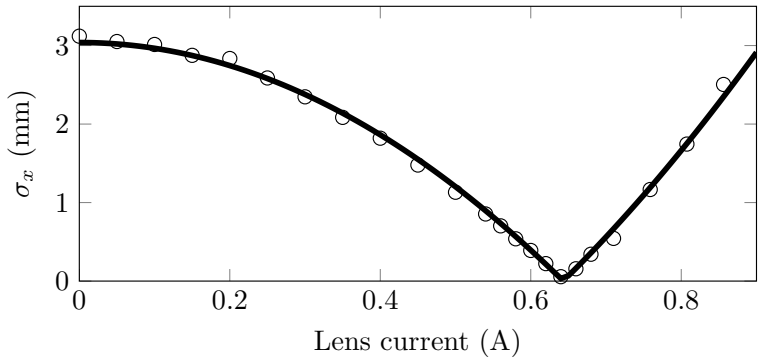}
\caption{The rms electron spot size $\sigma_{x}$ as measured on the detector for various magnetic lens currents.\label{waistscanimage}}
\end{figure}

We have repeated our previous measurements\cite{Engelen2013,Engelen2014,VanMourik2014a} with a  higher resolution electron detector (TVips TemCam-F216) allowing for more accurate measurements. Figure~\ref{waistscanimage} shows the spot size as measured on the detector as a function of magnetic lens current together with our beam line model\cite{Engelen2014,Engelen2013} fit which is used to determine the transverse beam emittance. These measurements result in a rms transverse normalized emittance of $\hat{\epsilon}_{x}= 1.5 \pm 0.1$~nm$\cdot$rad for $\lambda=490~$nm and $F=0.813~$MV/m. Using the value of the independently measured source size $\sigma_{x}=35~\mu$m\cite{Engelen2014}, this emittance corresponds to an effective transverse source temperature $T_{x}=10~$K confirming our earlier results\cite{Engelen2013,Engelen2014}.

\subsection{Longitudinal phase space}

The normalized longitudinal beam emittance of the ultracold electron source is, analogous to the transverse beam emittance (see Eq.~(\ref{emittancetrans})), described by:

\begin{equation}\hat{\epsilon}_{z}=\frac{\sigma_{z}\sigma_{p_{z}}}{mc}=\sigma_{z}\sqrt{\frac{k_{b}T_{z}}{m c^{2}}}.\label{emittance1}\end{equation}

with $\sigma_{p_{z}}$ the longitudinal rms momentum spread, $\sigma_{z}$ the rms size of the ionization volume in the acceleration ($\hat{z}$) direction and $T_{z}$ the effective longitudinal source temperature. In the most favorable case $T_{x}=10~$K, as measured, and $T_{z}=100~$K determined by the bandwidth ($\sigma_{\lambda}=4~$nm) of the ionization laser. This results in a normalized rms longitudinal emittance of $\hat{\epsilon_{z}}=4~$nm$\cdot$rad.
The longitudinal beam emittance is better known as the product of the rms pulse duration $\tau_{w}$ in a waist and the rms energy spread $\sigma_{U}$, 

\begin{equation}\hat{\epsilon}_{z}=\frac{\tau_{w} \sigma_{U}}{mc},\label{emittance2}\end{equation}

resulting in $\hat{\epsilon_{z}}\cdot mc=7~$ps$\cdot$eV.

The longitudinal beam emittance is an important parameter since it determines to what extent we can compress our electron pulse for a given energy spread, see Eq.~(\ref{emittance2}). Using Eq.~(\ref{relenergylasersize}),(\ref{emittance1}), (\ref{emittance2}) and $U=eFd_{acc}$ we can show that the rms pulse length in the longitudinal waist (See Fig.~\ref{accelerator}c) is given by

\begin{equation}\tau_{w}=\frac{\sqrt{mk_{b}T_{z}}}{eF}.\label{bla2145}\end{equation}

with $e$ the elementary charge. For $T_{z}=100$~K and $F=0.813$~MV/m we thus find $\tau_{w}=270$~fs. Strictly speaking, Eq.~(\ref{emittance2}) only holds when all electrons are created instantaneously. If the electrons are created over a time span $\tau_{ion}$, the longitudinal beam emittance becomes

\begin{equation}\hat{\epsilon}_{z}=\frac{\sigma_{U}}{mc} \sqrt{\tau_{w}^{2}+\tau_{ion}^2}.\label{bla1245}\end{equation}

This equation shows that the longitudinal beam emittance is influenced by the duration of the ionization process $\tau_{ion}$, which will be treated in the next section. 

Generally, the pulse length of the freely propagating bunch is determined by four processes: first the duration of the ionization laser pulse $\tau_{l}$; second the time $\tau_{ion}$ it takes an electron to escape the rubidium potential; Third, electron pulse lengthening due to the beam energy spread $\sigma_{U}$, as illustrated in Fig.~\ref{accelerator}; Fourth, the pulse lengthening due to space charge forces in the electron bunch. The latter process has been avoided in this work by using low charge densities.

The rms temporal bunch length $\tau_{U}$ due to the energy spread at a distance $z$ behind the MOT is given by,

\begin{equation}\tau_{U}\cong\sqrt{\frac{m\sigma^{2}_{ion}}{2U}}\left(1-\frac{z-d_{acc}}{2~d_{acc}}\right)\label{totalpulselength}.\end{equation}

Here we approximated the accelerating field by an uniform electric field $\vec{F}=F \hat{z}$ that extends up to $z=d_{acc}$ and is zero for $z > d_{acc}$. This approximation holds for positions $z-d_{acc} \gg a$, with $a$ the size of the aperture in the anode. This equation shows that there is a longitudinal focus at $z=3~d_{acc}$. Here the fast electrons created in the back of the ionization volume catch up with the slower electrons created in the front, as illustrated in Fig.~\ref{accelerator}.

\subsection{Ionization process}\label{simionmodel}

To predict the ionization time constant $\tau_{ion}$ we make use of a classical model of the ionization process. This model\cite{Engelen2014b,Engelen2013,Engelen2014,Engelen} has previously been used successfully to predict the transverse beam quality.

In the model describing the rubidium atom there are no closed electron orbits\cite{Engelen,Engelen2014b}, which means that all electrons with an excess energy $E_{exc}$ (Eq.~(\ref{excessenergy})) larger than zero will eventually leave the ion. The particle trajectories are calculated with the General Particle Tracer code\cite{GPT}. The electrons are started with a velocity directed radially outwards under an angle $\theta$ with the acceleration field. All simulation parameters are described in Ref\cite{Engelen2014b,Engelen}. From the simulated trajectories, the arrival time of the electrons is calculated at $z = 10~\mu$m, where the ion potential is negligible.

This is done for various starting angles and excess energies. To calculate the electron pulse shape for femtosecond laser ionization we have to convolve the simulation results with the broadband laser spectrum, with an rms width $\sigma_{\lambda}=4$~nm, and with the emission angle distribution associated with the polarization of the laser field. For polarization parallel to the acceleration field we assume an angular distribution $\sim \cos^2(\theta)$ and for perpendicular polarization we assume an angular distribution $\sim \sin^2(\theta)$.

\begin{figure}[htb!]
\includegraphics[width=14cm]{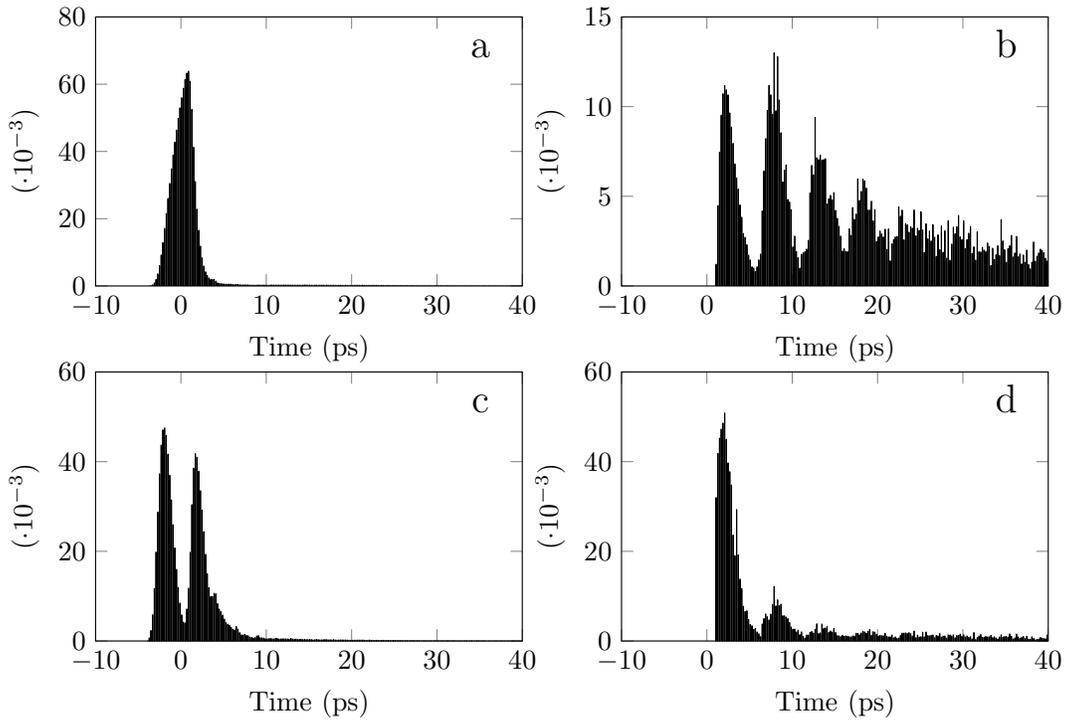}

\caption{\label{arrivaltimeraw} Normalized temporal electron bunch charge distribution. Images (a) and (b) ((c) and (d)) show the charge distribution for a laser polarization perpendicular (parallel) to the acceleration field. The Figures (a) and (c) ((b) and (d)) show the pulse shape for positive (negative) average excess energy for $\lambda=470$ ($489.8)~$nm.}
\end{figure}

The arrival-time distribution for $F=0.237$~MV/m, when using a femtosecond ionization laser pulse is depicted in Fig.~\ref{arrivaltimeraw} for ionization laser wavelengths $\lambda=470$ and $\lambda=489.8~$nm and ionization laser polarizations both parallel and perpendicular to the acceleration field. Figure~\ref{arrivaltimeraw}a and b show the temporal electron bunch charge distribution for a laser polarization perpendicular to the acceleration direction. 
Figure~\ref{arrivaltimeraw}a shows an electron pulse for a center wavelength of the ionization laser well above ($\lambda=470~$nm) the ionization threshold, resulting in a relatively large excess energy and a fast pulse. Figure~\ref{arrivaltimeraw}b shows an electron pulse for a center photon energy of the ionization laser pulse below ($\lambda=489.8$) the ionization threshold; the zero excess energy wavelength is at $\lambda=486~$nm. Electrons created with negative excess energies cannot escape (in the classical model) which means that we are effectively narrowing the bandwidth of the broadband laser pulse. This results in a train of electron pulses leaving the atom\cite{Robicheaux1997,Bordas2003,Lankhuijzen1996a,Engelen}.

Figure~\ref{arrivaltimeraw}c and d show the temporal electron bunch charge distribution for a laser polarization parallel to the acceleration direction. Figure~\ref{arrivaltimeraw}c shows a fast pulse which is split in two. The first pulse is due to the electrons that immediately exit the potential; the second pulse is due to the electrons that are launched in the uphill direction (see Fig.~\ref{ionschemerb}d) and subsequently first make a round trip inside the potential before leaving. Figure~\ref{arrivaltimeraw}d results in an electron pulse train, similar to Fig.~\ref{arrivaltimeraw}b. The ratio of the intensity of the peaks belonging to the second and third electron pulse compared to the first pulse are much smaller than that for perpendicular polarization. This is caused by the fact that there is a maximum ejection angle $\theta_{c}$\cite{Engelen2014b,Engelen} for a given $\lambda$ and $F$, so that for $\parallel$ polarization almost half of the initial angles $\theta \leq \theta_{c}$.

In our experiments the ionization laser pulse length $\tau_{l}\approx 100~$fs; the pulse length contribution due to the laser pulse length can therefore be neglected as $\tau_{ion}$ is much larger.

\section{Experimental}\label{sectionexperi}

\begin{figure}[htb!]
\centering
\includegraphics[width=14cm]{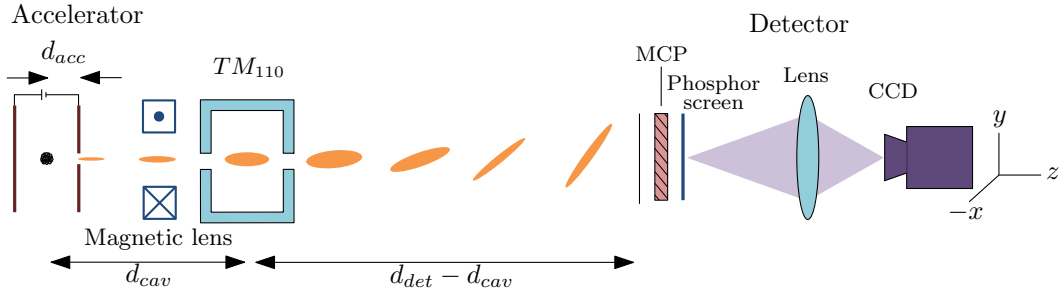}
\caption{\label{beamline}Schematic representation of the beam-line. First the electrons are accelerated in a DC electric field. A set of steering coils and a magnetic solenoid lens is used to steer and focus the electron beam. The TM$_{110}$~RF cavity deflects the electron pulse by an amount that depends on the arrival time. The electrons are detected by a dual micro-channel plate in combination with a phosphor screen. The light emanating from this screen is imaged onto a CCD camera.}
\end{figure}

Figure~\ref{beamline} shows a schematic representation of the entire beam line. The electrons are created at the center of a DC accelerating structure\cite{Taban2008}. The accelerated electrons pass a set of steering coils and a magnetic solenoid lens schematically indicated by a single magnetic lens. These electron optical elements allow us to control the beam position and size. Before the electron beam reaches the detector, which consists of a dual micro-channel plate (MCP) and a phosphor screen, it passes through a $3$~GHz RF deflecting cavity which will be used to measure the electron bunch lengths. Note that the detector used for the pulse length measurements (dual MCP) is different from the detector that has been used for the waist scan measurements described in section~\ref{waistscansecti}.

\subsection{RF cavity}

The electric and magnetic fields present in an RF cavity operated in the TM$_{110}$ mode, exert a mainly transverse force on the electrons whose strength depends on the RF phase at the moment they pass through. The electrons in a bunch of finite length, shorter than half an RF oscillation period will therefore acquire a transverse momentum kick while traveling through the cavity whose magnitude depends on their arrival time. The bunch will then be streaked across a detector downstream, as illustrated in Fig.~\ref{beamline}. To enable measurements of sub-picosecond bunch lengths the cavity needs to be operated in the GHz regime. A $3~$GHz streak cavity was recently developed in our group\cite{Lassise2012a,Lassise2012}, optimized for a $30$~keV beam. We will use this cavity to probe the length of the electron bunches extracted from the MOT.

The phase of the electromagnetic fields inside the RF cavity is synchronized\cite{Brussaard2013} with few $100~$fs accuracy to the femtosecond laser pulse which ionizes the rubidium atoms,  guaranteeing that the center of every electron pulse experiences nearly the same electromagnetic field every time it passes through the cavity. For a pillbox cavity with small entrance and exit holes at $z=\pm \frac{L_{cav}}{2}$ and transverse positions close to the cavity axis, i.e. $x,y \ll \frac{c}{\omega}$, the magnetic field can be approximated by:

\begin{equation}\vec{B}(x,y,z,t) \approx B_{0}~\sin(\phi+\omega t)~\hat{x}\label{lol2},\end{equation}

with $\omega$ the angular frequency, $\phi$ a phase offset and $B_{0}$ the amplitude of the oscilating magnetic field inside the cavity. We can show that the rms length of the resulting streak on the detector is given by:

\begin{equation} \sigma_{screen}^{2}=\sigma_{off}^{2} + \left(2 \omega_{c} \sigma_{t}~(d_{det}-d_{cav})~\sin(\zeta)\cos(\phi)\right)^{2}\label{sigmaanastreak},\end{equation}

while the average deflection angle of the electron pulse is given by

\begin{equation} \frac{\Delta v_{y}}{v_{z}}=\frac{2\omega_{c}}{\omega}\sin(\zeta)\sin(\phi)\label{avgdeflection},\end{equation}

where $\zeta\equiv\frac{\omega L_{cav}}{2 v_{z}}$, $\sigma_{off}$ is the transverse rms beam size when the cavity is turned off and $\omega_{c}=\frac{eB_{0}}{m}$ the cyclotron frequency. To maximize the streak length the bunch has to stay either one ($\zeta=\frac{\pi}{2}$) or three half periods ($\zeta=\frac{3\pi}{2}$) inside the cavity. The cavity length $L_{cav}=16.7~$mm was optimized to $\zeta=\frac{\pi}{2}$ for a $30$~keV beam\cite{Lassise2012a,Lassise2012}. Since the energy in this setup is $\leq 20~$keV, the electron beam energy is fixed at $U=3.2~$keV, resulting in $\zeta=\frac{3\pi}{2}$.

The $3$~GHz deflecting cavity is positioned at a distance $d_{cav}=0.68$~m from the magneto-optical trap while the detector is at a distance $d_{det}=1.9~$m. The pulse length $\sigma_{t}$ at the position of the cavity is dominated by the energy spread of the electron beam. From Eq.~(\ref{totalpulselength}) we can estimate this $\sigma_{\tau}\approx 18$~ps. Here we have assumed that the ionization laser beam is at its diffraction limit ($\sigma_{ion}=30~\mu$m) and the pulse length $\tau_{l}\ll 1$~ps. Furthermore we assume that the average beam energy is optimized for maximum streak ($\zeta=\frac{3\pi}{2}$) and that the duration of the ionization process $\tau_{ion} \lesssim 10$~ps, as discussed in section~\ref{simionmodel}. 
The pulse length measurements are calibrated by measuring the term $\frac{2 \omega_{c}}{\omega}\sin(\zeta)$ in Eq.~(\ref{avgdeflection}). This is done by scanning the phase $\phi$ while measuring the position on the screen $\frac{\Delta v_{y}}{v_{z}} (d_{det}-d_{cav})$. The phase $\phi$ can be scanned with a phase shifter (Mini-Circuits JSPHS-446) by applying a relative voltage $v_{phase}=\frac{V_{phase}}{V_{max}}$ with $0\leq v_{phase} \leq 1$.

\subsection{Ionization laser}

The femtosecond laser consists of an amplified Ti:Sapphire laser system (Coherent Legend Elite) that pumps an optical parametric amplifier (Coherent OPerA Solo) generating tunable $480$~nm femtosecond pulses. The Ti:Sapph system produces $800$~nm $35$~fs pulses with an energy of $2.5~$mJ per pulse at a repetition frequency of $1$~kHz. The $480$~nm laser pulse length $\tau_{l}$ at the MOT is estimated to be $100$~fs. The pulse energy of the ionization laser increases as a function of wavelength, from $\sim 75~\mu$J for $\lambda=470~$nm to $\sim 150~\mu$J for $\lambda=490~$nm.


\section{Results}\label{secresultsanddisc}

\begin{figure}[htb!]
\centering
\includegraphics[width=12cm]{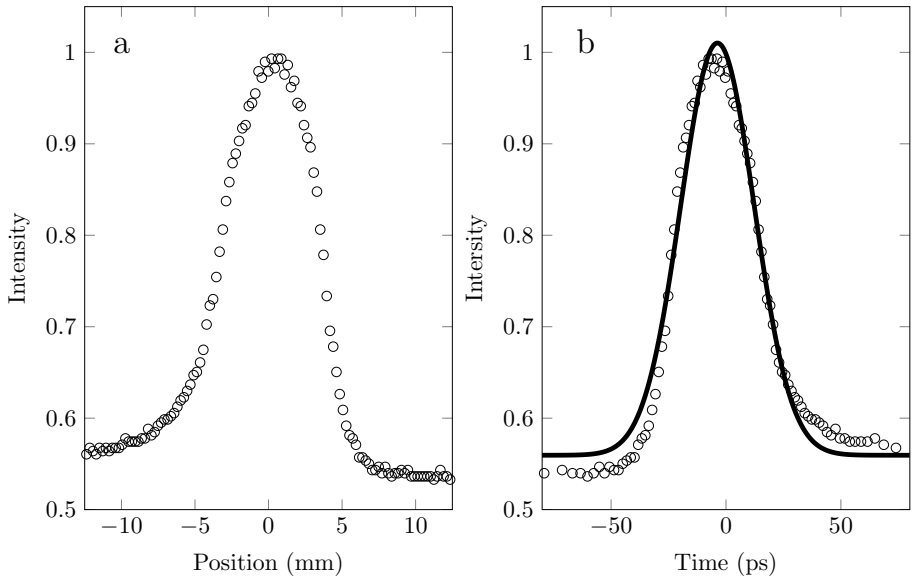}
\caption{\label{spatialconvtimerpoal}(a) Example of an electron beam profile as measured on the detector. (b) Electron beam profile in time domain plus gaussian fit (curve) through the data (dots) to determine the rms electron pulse length.}
\end{figure}

An example of an streak as measured on the detector is depicted in  Fig.~\ref{spatialconvtimerpoal}a. This figure is recorded with an ionization laser wavelength $\lambda=483$~nm and using the $(480-480)$ ionization scheme. Every streak measurement consists of $\sim 10^{3}$ electron pulses. For every wavelength $\lambda$ of the ionization laser, the phase $\phi$ of the RF cavity was scanned over one entire period while the electron beam was imaged by the detector. 

\begin{figure}[htb!]
\centering
\includegraphics[width=11cm]{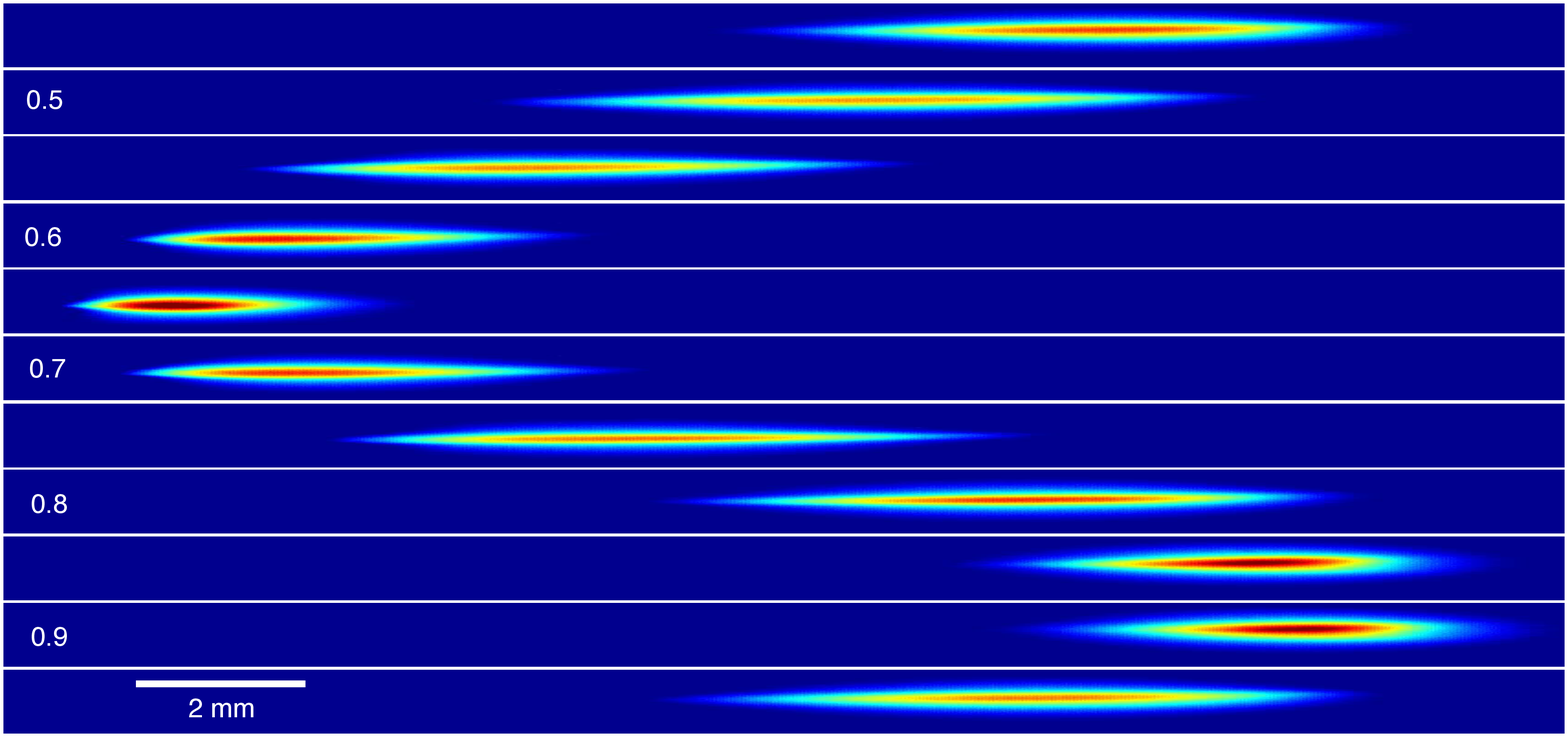}
\caption{\label{colorplot} False color plot of the electron beam as measured on the detector for relative phase voltages $v_{phase}$ ranging from $0.45$ (top) to $0.95$ (bottom) in steps of $0.05$.}
\end{figure}

Figure~\ref{colorplot} shows a false color plot of the electron spot as measured on the detector as function of the relative phase voltage $v_{phase}$. The figure clearly shows that the electron spot is swept across the detector, as is predicted by Eq.~(\ref{avgdeflection}).

\begin{figure}[htb!]
\centering
\includegraphics[width=12cm]{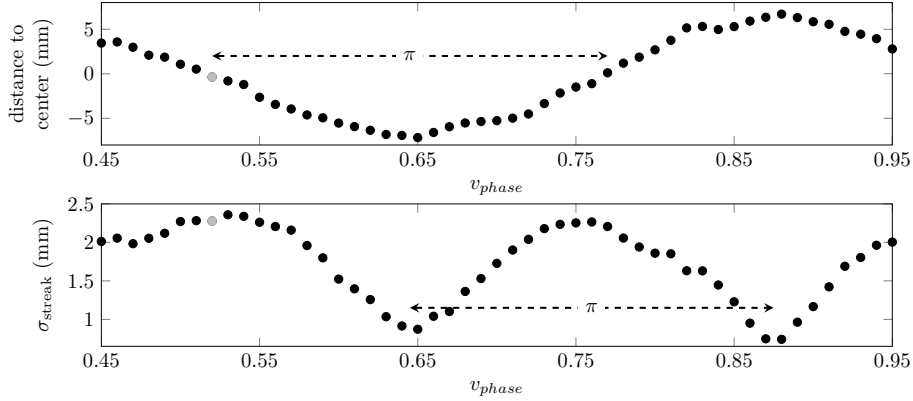}
\caption{\label{posonstreak} Top: Distance to the center position of the streak as function of relative phase shifter voltages $v_{phase}$ ranging from $0.45$ to $0.95$. Bottom: The rms size of the electron spot as a function of the cavity phase. The grey dot indicates the measurement shown in Fig.~\ref{spatialconvtimerpoal}.}
\end{figure}

The position of the electron pulse with respect to the center of the streak for relative phase shifter voltages $v_{phase}$ ranging from $0.45$ to $0.95$ is depicted in the top plot of Fig.~\ref{posonstreak}. This figure nicely shows that scanning the phase of the RF cavity will shift the position of the electron spot across the detector, as shown in Fig.~\ref{colorplot}. The position of the example electron spot (see Fig.~\ref{spatialconvtimerpoal}a) is indicated by the grey dot.

The bottom plot of Fig.~\ref{posonstreak} shows the rms size of the electron spot, which was obtained by fitting with a gaussian function. This shows that the cavity is most sensitive for arrival time spread when the electron pulse is on the center of the streak, as predicted by Eq.~(\ref{sigmaanastreak}) which is clearly visible in Fig.~\ref{colorplot}. Knowing the total length of the streak (used to determine $\frac{2 \omega_{c}}{\omega}\sin(\zeta)$) we can calibrate the time axis separately for each wavelength. The right plot of Fig.~\ref{spatialconvtimerpoal} shows the electron pulse in the time domain together with a gaussian fit resulting in a rms pulse length of $15~$ps.

In the next section we first present the streak data of the direct photoionization scheme ($480-480$) and the just-above-threshold photoionization scheme ($780-480$). Subsequently we show that we can make a pulse train by shaping the ionization laser beam profile. Finally we present pulse length measurements of slowly ionizing Rydberg states using the just-below-threshold photoionization scheme ($780-480-$Ry).

\subsection{Direct photoionization ($480-480$)}

\begin{figure}[htb!]
\centering
\includegraphics[width=12cm]{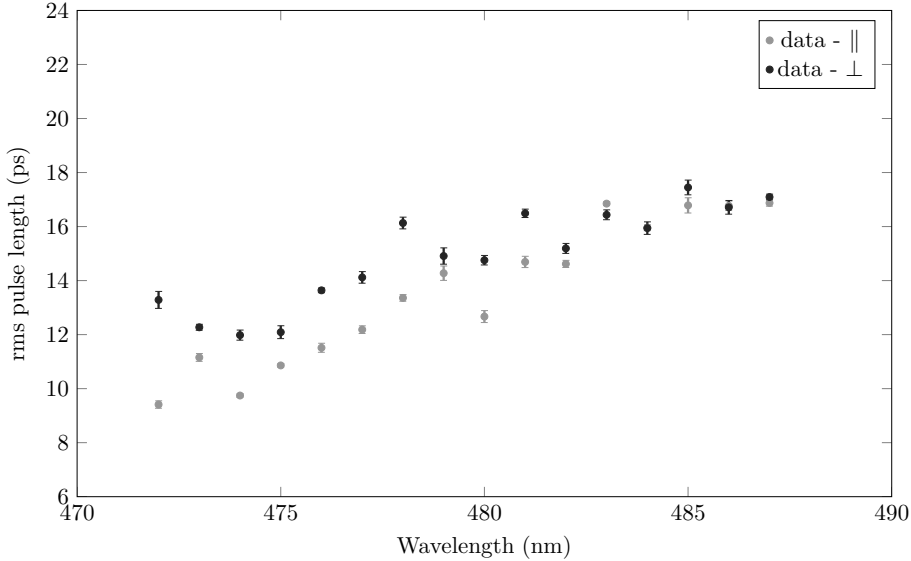}
\caption{\label{pulse2blue}The rms pulse length as a function of ionization laser wavelength for direct photoionization $(480-480)$ for laser polarization perpendicular (black) and parallel (grey) to the acceleration field. }
\end{figure}

Figure~\ref{pulse2blue} shows the rms pulse length of electron pulses created by direct ionization from the rubidium ground state (Fig.~\ref{ionschemerb}a). The measurement has been done both for laser polarization parallel and perpendicular to the acceleration field.
The rms pulse lengths measured here are shorter than can be explained by the diffraction-limited rms size of the ionization laser, which is in agreement with the fact that direct ionization scales with the square of the intensity of the laser field effectively narrowing the rms size of the ionization volume by a factor of $\frac{1}{\sqrt{2}}$. A diffraction limited ionization laser beam $\sigma_{ion}=30~\mu$m should result in $18/\sqrt{2}\approx 13~$ps, which is confirmed by the measurement presented in Fig~\ref{pulse2blue}. We note that the amount of electrons per pulse is smaller for $\parallel$ polarization in contrast to $\perp$ polarization. We also find that the measured rms pulse lengths are shorter for $\parallel$ than for $\perp$ polarization and that the pulse length increases with the ionization wavelength. These experimental findings are not yet fully understood and require further investigation, which is outside the scope of this paper.

The $\sim 1~$ps variation in the data points in Fig.~\ref{pulse2blue} can be explained by a $\sim 1~\mu$m pointing instability of the ionization laser beam.

\subsection{Just-above-threshold photoionization ($780-480$)}

\begin{figure}[htb!]
\centering
\includegraphics[width=12cm]{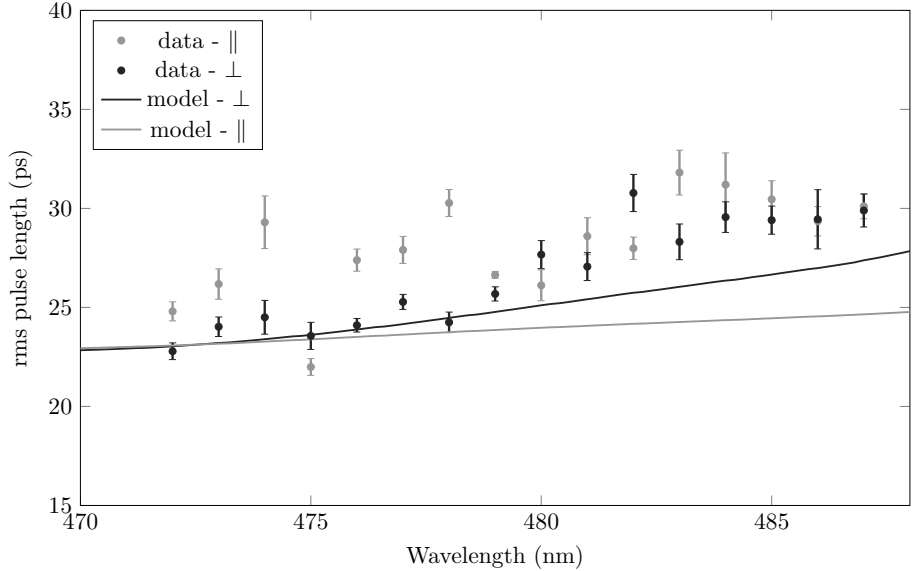}
\caption{\label{twostepioniz}The rms pulse length as a function of laser wavelength for near-threshold photoionization ($780-480$) of a rubidium atom for laser polarization perpendicular (grey) and parallel (black) to the acceleration field. The model is indicated by the solid curves.}
\end{figure}

Figure~\ref{twostepioniz} shows the measured rms pulse length of an electron pulse created by just-above-threshold ionization of excited rubidium atoms (Fig.~\ref{ionschemerb}b). 

The pulse length at the position of the cavity is predominantly determined by the energy spread of the electron bunch, see Eq.(\ref{totalpulselength}). Convolving the temporal electron pulse distributions, see Fig.~\ref{arrivaltimeraw}, with a gaussian energy spread given by a gaussian ionization laser beam with a rms width of $\sigma_{ion}=32~\mu$m we can calculate the expected pulse length for various ionization laser wavelengths and polarizations. The results are represented by the solid lines in Fig.~\ref{twostepioniz}.

\begin{figure}[htbp]
\centering
\includegraphics[width=12cm]{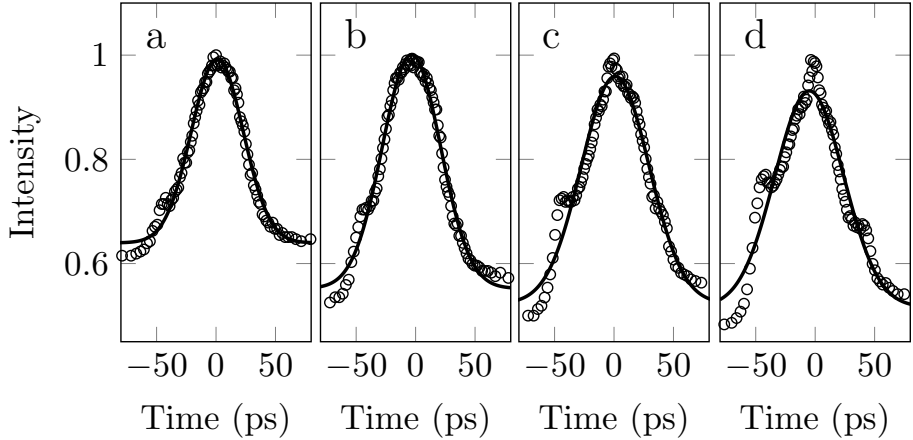}
\caption{\label{electronpulseshape2step} Electron pulse in the time domain for increasing ionization laser wavelengths. From left to right, $\lambda=472, 479, 480, 487~$nm. The curve indicates a gaussian fit through the data which has been used to determine the rms pulse duration.}
\end{figure}

We see that the measured rms pulse length is in agreement with the pulse length determined by the energy spread. We also see that the rms pulse length increases as a function of wavelength but the increase is relatively small with respect to the magnitude of the pulse lengths, as expected from the simulations. The data shows a stronger growth than expected. Similar to the direct ionization scheme ($480-480$) the measurement resolution is limited due to pointing instabilities of the ionization laser beam. Additionally, the gaussian fits are less reliable due to deviation from perfect gaussian behavior, as will be discussed below, see Fig.~\ref{electronpulseshape2step}.

The electron pulse shapes for various ionization laser wavelengths are depicted in Fig.~\ref{electronpulseshape2step}, together with their gaussian fits which were used to determine the rms pulse lengths presented in Fig.~\ref{twostepioniz}. Figures b,c and d show sharp features around $\pm40$~ps. These are probably due to a deviation from a perfect gaussian ionization laser beam profile. The features do not change position as a function of wavelength and are too sharp to be explained by a pulse train emitted by the rubidium atom (see Section~\ref{simionmodel}) since this temporal information is washed out due to the energy spread of the beam, resulting in features with at least an rms width of $20$~ps. The pulse train predicted by the simulations (See Fig.~\ref{arrivaltimeraw}) cannot be measured since the time difference between the pulses is smaller than the pulse broadening due to the energy spread.

Figure~\ref{electronpulseshape2step}c and d show a gaussian arrival time distribution with a very sharp peak in the center. This peak can be attributed to slowly ionizing Rydberg atoms. This effect will be discussed in Section~\ref{autorydbergscet}.

These are measurements of \emph{both} ultracold \emph{and} ultrafast electron bunches containing $\sim 10^{3}$ electrons per pulse. The waist scan presented in Section~\ref{waistscansecti} has been performed on the same beam with the cavity retracted from the beam line with $\lambda=490~$nm and $F=0.814~$MV/m. This shows that we can produce electron pulses containing $\sim 10^{3}$ electrons with an rms width of $\sim 25$ picosecond and a normalized rms transverse emittance of $1.5\pm0.1$~nm$\cdot$rad.

\subsection{Pulse shaping}

\begin{figure}[htb!]
\centering
\includegraphics[width=12cm]{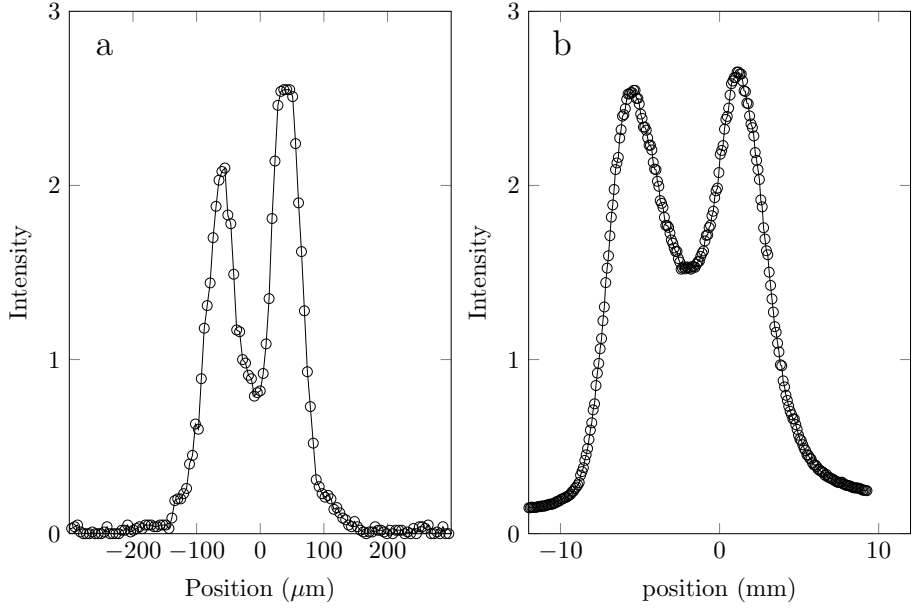}
\caption{\label{donutstreak}(a) Intensity profile of the femtosecond ionization laser beam as measured on a CCD camera. (b) Electron pulse as measured on the detector.}
\end{figure}

We will now show how transversely shaping the ionization laser beam profile allows us to temporally shape an electron pulse. Figure~\ref{donutstreak}a shows an example in which the ionization laser profile was shaped such that the distance between the two peaks is $90~\mu$m. This distribution will lead to a similarly shaped longitudinal energy distribution and thus in a similarly shaped arrival time distribution at the cavity. Using Eq.~(\ref{totalpulselength}) we can estimate that the temporal electron pulse length at the position of the cavity is $\sim 65~$ps. Figure~\ref{donutstreak}b shows the streak as measured on the MCP detector. The peaks in this figure are $\sim 6.5~$mm apart. A full streak is equal to $14~$mm, see top plot of Fig.~\ref{posonstreak}, which is equivalent to half a RF period $~\sim 160~$ps. This means that the measured pulse length $\sim 75~$ps, roughly in agreement with the expected pulse length. We thus show that the intensity distribution of the ionization laser profile is indeed imprinted on the temporal charge distribution we measure.

This opens the possibility to produce well defined pulse trains by shaping the intensity profile of the ionization laser.

\subsection{Just-below-threshold photoionization ($780-480-$Ry)}\label{autorydbergscet}

We have seen the first evidence of slowly decaying Rydberg atoms in Fig.~\ref{electronpulseshape2step}c and d. Here we will discuss how these slowly ionizing Rydberg atoms are formed when part of the ionization laser spectrum is below the ionization threshold. 

The RF power to the cavity is switched on $10~\mu$s before the ionization laser pulse reaches the MOT, to make sure that the electromagnetic fields inside the cavity are stable. In all the above mentioned measurements the RF power to the cavity was switched off $1~\mu$s after the $480~$nm ionization laser pulse had reached the MOT. This was done to prevent unnecessary heating of the RF cavity thus reducing phase instabilities. By this time all the fast electrons have reached the detector (Time of flight is $\sim~50~$ns). 

\begin{figure}[htb!]
\centering
\includegraphics{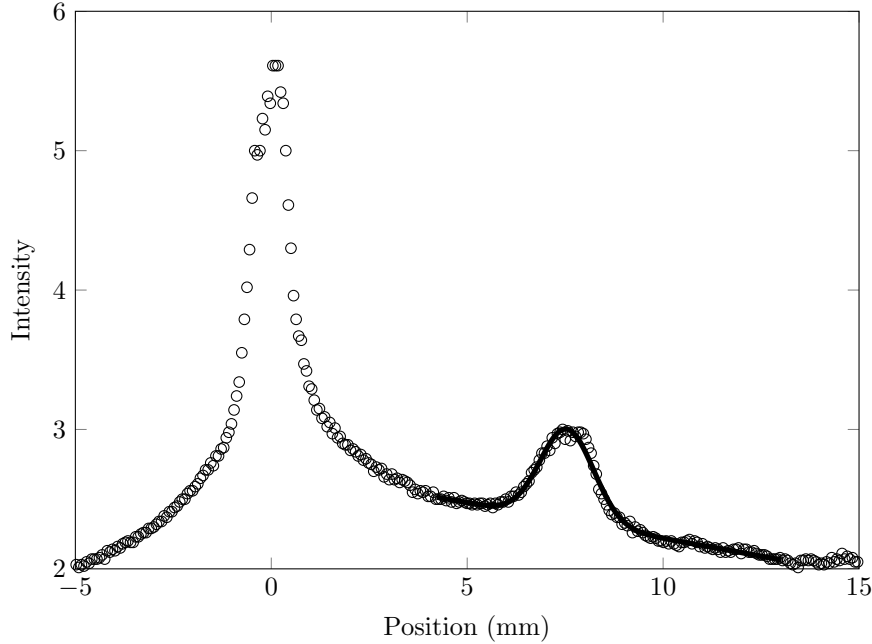}
\caption{\label{rydbergpeaks}Streak of the electron beam as measured on the detector. The left peak (between $-5$ and $5~$mm) indicates the electrons that are streaked by the cavity and the right peak (between $5$ and $10~$mm) shows the electrons that have passed the cavity after the RF power was switched off. The solid curve indicates a gaussian fit through the data.}
\end{figure}

Due to the relatively low quality factor of the cavity the time it takes for the fields inside the cavity to build up and decay is $\sim100$~ns. Electrons traversing the cavity after the RF power was switched off will not be deflected and will therefore pass right through. Varying the time at which the RF power is switched off allows us to measure the fraction of electrons that pass the cavity at timescales $\sim \mu$s after the ionization laser pulse.

Figure~\ref{rydbergpeaks} shows the streak of an electron pulse when the RF power was switched off $1~\mu$s after the laser pulse had reached the MOT. The left peak (between $-5$ and $5~$mm) shows the fast electron pulse that is streaked by the cavity; the right peak (between $5$ and $10~$mm) shows electrons that have passed the cavity after the RF power was switched off. The solid line indicates a gaussian fit through the undisturbed electron peak.

\begin{figure}[htb!]
\centering
\includegraphics[width=15cm]{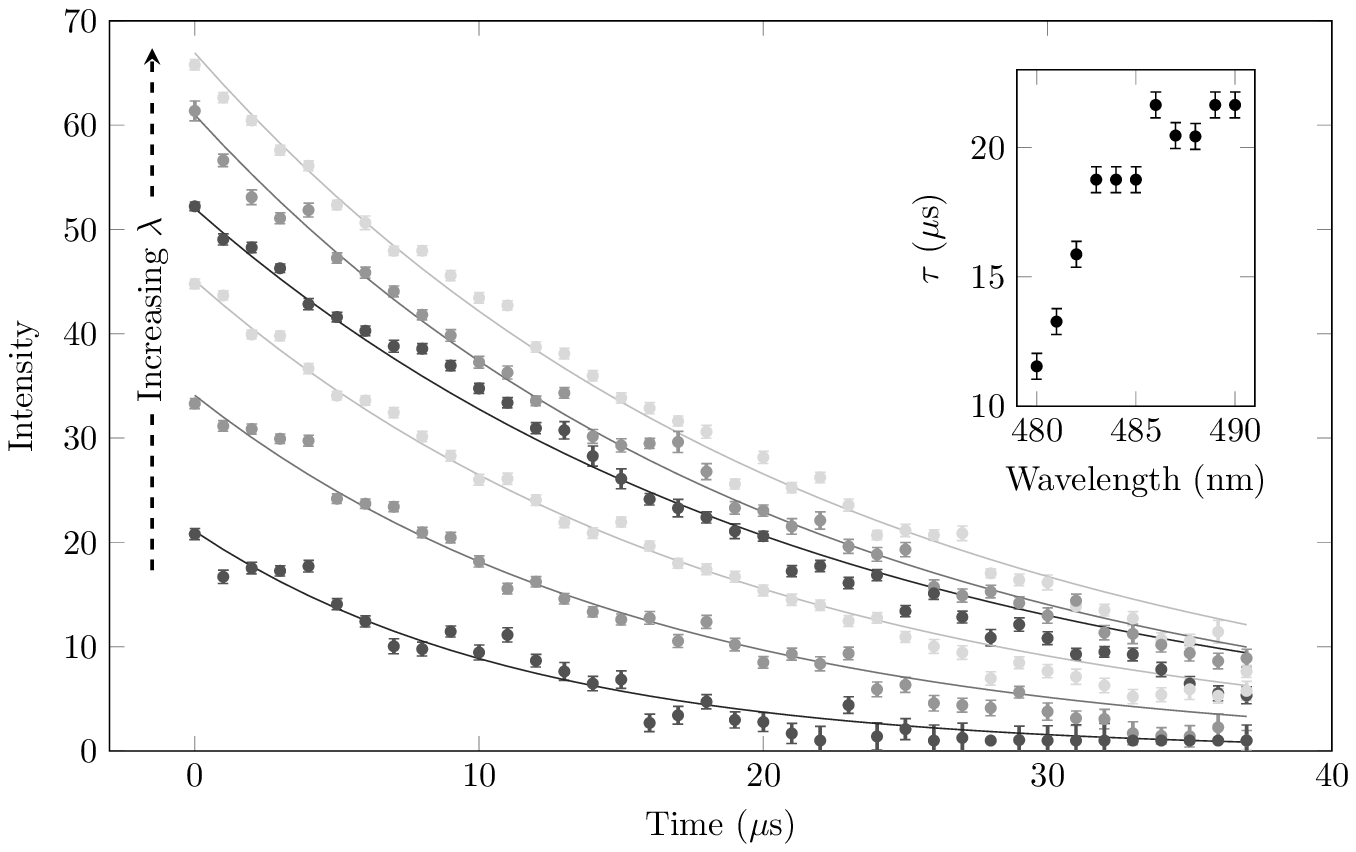}
\caption{\label{rydbergpeaksexpdecay}The decay of the population of excited Rydberg atoms in the MOT versus time for various ionization laser wavelengths (from bottom to top; $\lambda=480$,$483$,$485$,$487$,$489$,$490~$nm). Increasing the ionization laser wavelength increases the number of excited Rydberg atoms which also tend to decay slower, as shown by the inset plot.}
\end{figure}

We have measured the streak as depicted in Fig.~\ref{rydbergpeaks} as a function of the time after which the RF power was switched off. Figure~\ref{rydbergpeaksexpdecay} shows the intensity of the peak going straight through versus the time the RF power was on after the ionization laser pulse hit the MOT, for various ionization laser wavelengths. The lowest lying data points in Fig.~\ref{rydbergpeaksexpdecay} represent a center ionization laser wavelength close to the ionization threshold ($\lambda=480~$nm). Increasing the ionization laser wavelength increases the intensity of the Rydberg signal. The intensity of the peaks shows an exponential decay and are fitted with $\exp(-t/\tau)$ indicated by the solid curves. The decay constant $\tau$ as function of wavelength is depicted in the inset plot of Fig.~\ref{rydbergpeaksexpdecay}. The observed decay times are in agreement with earlier observations\cite{Robinson2000}. Note that the larger the wavelength of the ionization laser the more Rydberg atoms are created, as expected. In addition the time constant of the ionization process increases when the ionization laser wavelength is increased. This is attributed to the fact that lower-lying Rydberg states take longer to ionize. 



\section{Conclusions and outlook}\label{outlookconc}

An RF cavity has been used to measure the pulse length of electron bunches produced by femtosecond laser photoionization of a laser-cooled and trapped ultracold atomic gas. Pulse lengths have been measured in three photoionization regimes: direct photoionization using only the femtosecond laser, two-step just-above-threshold photoionization, and two-step just-below-threshold photoionization. Both direct and just-above-threshold ionization produce ultrashort electron pulses with rms pulse durations ranging from $10$ to $30$~ps. Direct ionization produces few ($\sim10^{2}$) and hot ($\sim10^{4}~$K) electrons while just-above-threshold ionization produces many ($\sim10^{3}$) and ultracold ($\sim10~$K) electrons with a normalized transverse emittance $\hat{\epsilon}_{x}=1.5\pm0.1$~nm$\cdot$rad. The measured pulse lengths can be explained by the energy spread associated with the length of the ionization volume. These bunches are sufficiently short to be compressed to sub-ps pulse lengths using established RF compression techniques.
Just-below-threshold ionization produces a Rydberg gas which ionizes on $\sim \mu$s timescales.

The method described here can be used to investigate the effect of space charge forces which is the main limitation of the achievable temporal resolution in single-shot UED. Measuring the pulse length in the longitudinal waist should allow the direct measurement of the influence of space charge forces on the longitudinal phase space distribution.

The temporal distribution of the predicted train of electron pulses leaving a rubidium atom could also be investigated by measuring the pulse length in the longitudinal waist.


\begin{acknowledgments}

This research is supported by the Institute of Complex Molecular Systems (ICMS) at Eindhoven University of Technology. The authors would like to thank Eddy Rietman and Harry van Doorn for expert technical assistance.

\end{acknowledgments}


%

\end{document}